\documentclass[10pt,a4paper,aps,prd,twocolumn,preprintnumbers]{revtex4}

\usepackage{epsfig}
\usepackage{amsmath}
\usepackage{amssymb}
\usepackage{amsfonts}
\usepackage{graphicx}
\usepackage{graphics,subfigure}
\usepackage{float}
\usepackage{booktabs}
\usepackage{epsfig}
\usepackage{color}
\usepackage{geometry}
\usepackage{rotating}
\usepackage{psfrag}
\usepackage{supertabular}

\newcommand{\newc}{\newcommand}

\newc{\eps}{\epsilon}
\newc{\lamp}{\lambda^{\prime}}
\newc{\Lam}{\Lambda}
\newc{\ra}{\rightarrow}
\newc{\lra}{\leftrightarrow}
\newc{\wtilde}{\widetilde}
\newc{\ie}{{\it i.e.}}
\newc{\eg}{{\it e.g.}}
\newc{\rpv}{\not\!\! M_p}
\newc{\lsim}{\stackrel{<}{\sim}}
\newc{\beq}{\begin{equation}}
\newc{\eeq}{\end{equation}}
\newc{\beqn}{\begin{eqnarray}}
\newc{\eeqn}{\end{eqnarray}}
\newc{\PLB}{\emph{Phys.Lett.}{\bf{B}}}
\newc{\NPB}{\emph{Nucl.Phys.}{\bf{B}}}
\newc{\mcal}{\mathcal}
\newc{\bsym}{\boldsymbol}
\newc{\nonum}{\nonumber}
\newc{\tA}{\tilde{A}}
\newc{\tB}{\tilde{B}}
\newc{\ttau}{\tilde{\tau}}
\newc{\tAa}{\tilde{A}_\alpha}
\newc{\tAb}{\tilde{A}_\beta}
\newc{\tchi}{\tilde{\chi}}
\newc{\tl}{\tilde{\ell}}
\newc{\tnu}{\tilde{\nu}}
\newc{\tu}{\tilde{u}}
\newc{\td}{\tilde{d}}
\newc{\bu}{\bar u}
\newc{\bv}{\bar v}
\newc{\slashpt}{\slashed{\tau}}
\newc{\slashp}{\slashed{\nu}_\tau}
\newc{\slashq}{\tilde{\slashed{\tau}}}
\newc{\AmAA}{Re(\overline{A_l^\alpha A_{l'}^{\alpha'*}})}
\newc{\AmBB}{Re(\overline{B_m^\beta B_{m'}^{\beta'*}})}
\newc{\AmCC}{Re(\overline{C_n^\gamma C_{n'}^{\gamma'*}})}
\newc{\AmAB}{Re(\overline{A_l^\alpha B_{m}^{\beta*}})}
\newc{\AmAC}{Re(\overline{A_l^\alpha C_{n}^{\gamma*}})}
\newc{\AmBC}{Re(\overline{B_m^\beta C_{n}^{\gamma*}})}
\newc{\fquad}{\quad \quad \quad \quad \quad}
\definecolor{Red}{cmyk}{0,1,1,0}
\definecolor{luhn}{rgb}{1,0,0.5}
\definecolor{thor}{rgb}{0,0.6,1}
\definecolor{RubineRed}{cmyk}{0,1,0.13,0}
\newc{\colRR}{\color{RubineRed}}
\newc{\Red}{\color{Red}}

\renewcommand{\eqref}[1]{Eq.~(\ref{#1})}
\newc{\eqsref}[2]{Eqs.~(\ref{#1}),\,(\ref{#2})}
\newc{\figref}[1]{Fig.~\ref{#1}}

\newc{\lan}{\mathcal{L}}
\newc{\abs}[1]{\lvert#1\rvert}
\newc{\oas}{\mathcal{O}(\alpha_s^2)}
\newc{\rb}[1]{\raisebox{1.5ex}[-1.5ex]{#1}}

\newc{\ba}{\begin{array}}
\newc{\ea}{\end{array}}
\newc{\bea}{\begin{eqnarray}}
\newc{\eea}{\end{eqnarray}}
\newc{\beastar}{\begin{eqnarray*}}
\newc{\eeastar}{\end{eqnarray*}}
\newc{\mymed}{\vspace{0.19cm}}

\newbox\charbox
\newbox\slabox
\def\s#1{{      
    \setbox\charbox=\hbox{$#1$}
    \setbox\slabox=\hbox{$/$}
    \dimen\charbox=\ht\slabox
    \advance\dimen\charbox by -\dp\slabox
    \advance\dimen\charbox by -\ht\charbox
    \advance\dimen\charbox by \dp\charbox
    \divide\dimen\charbox by 2
    \raise-\dimen\charbox\hbox to \wd\charbox{\hss/\hss}
    \llap{$#1$}
}}

\newc{\lampp}{\lambda^{\prime\prime}}
\newc{\BLam}{{\mathbf{\Lam}}}
\newc{\kap}{\kappa}

\def\eq{\begin{equation}}
\def\qe{\end{equation}}

\newc{\azero}{A_0}
\newc{\neutralino}{\tilde\chi^0}
\newc{\selectron}{\tilde e}
\newc{\stau}{\tilde\tau}
\newc{\smuon}{\tilde\mu}
\newc{\sgnmu}{\textrm{sgn}(\mu)}
\newc{\gev}{\,\mbox{GeV}}
\newc{\tev}{\mbox{~TeV}}
\newc{\pb}{\mbox{~pb}}
\newc{\gsim}{\stackrel{>}{\sim}}

\newc{\del}{\partial}
\newc{\veva}{\langle H_1\rangle}
\newc{\vevb}{\langle H_2\rangle}
\newc{\onehalf}{\textstyle \frac{1}{2} \displaystyle}
\newc{\onethird}{\textstyle \frac{1}{3} \displaystyle}
\newc{\mzero}{M_0}
\newc{\mhalf}{{M_{1/2}}}
\newc{\tanb}{\tan\beta}
\newc{\Psix}{{\mathrm{P}_{\!6}}}
\newc{\nPsix}{{\not\!\Psix}}
\newc{\nPsixU}{{\s{\mathrm P}_{\!6}}}
\newc{\sneutrino}{\tilde\nu}

\newc{\ovl}{\overline}

\newc{\ssup}{\tilde{u}}
\newc{\ssdown}{\tilde{d}}
\newc{\ssstrange}{\tilde{s}}
\newc{\sscharm}{\tilde{c}}
\newc{\sstop}{\tilde{t}}
\newc{\ssbottom}{\tilde{b}}

\newc{\sse}{\tilde{e}}
\newc{\ssmu}{\tilde{\mu}}
\newc{\sstau}{\tilde{\tau}}
\newc{\ssnue}{\tilde{\nu}_{e}}
\newc{\ssnumu}{{\tilde{\nu}_{\mu}}}
\newc{\ssnutau}{{\tilde{\nu}_{\tau}}}
\newc{\ssbnue}{\bar{\tilde{\nu}}_{e}}
\newc{\ssbnumu}{\bar{\tilde{\nu}}_{\mu}}
\newc{\ssbnutau}{\bar{\tilde{\nu}}_{\tau}}

\newc{\neut}{{\tilde\chi}^0}
\newc{\charge}{{\tilde\chi}}
\newc{\glu}{\tilde{g}}

\newc{\higgs}{h^0}
\newc{\Higgs}{H^0}
\newc{\Azero}{A^0}

\newc{\nue}{\nu_e}
\newc{\numu}{\nu_{\mu}}
\newc{\nutau}{\nu_{\tau}}
\newc{\bnue}{\bar{\nu_e}}
\newc{\bnumu}{\bar{\nu_{\mu}}}
\newc{\bnutau}{\bar{\nu_{\tau}}}

\newc{\Mgut}{M_X}
\newc{\softsusy}{\texttt{SOFTSUSY\;}}
\newc{\lam}{\lambda}

\def\lsim{\raise0.3ex\hbox{$\;<$\kern-0.75em\raise-1.1ex\hbox{$\sim\;$}}}
\def\gsim{\raise0.3ex\hbox{$\;>$\kern-0.75em\raise-1.1ex\hbox{$\sim\;$}}}
\begin{document}

\title{Single Slepton Production in association with a single Top Quark 
at the Tevatron and LHC}

\author{M.~A.~Bernhardt}
\email[]{markus@th.physik.uni-bonn.de}
\author{H.~K.~Dreiner}
\email[]{dreiner@th.physik.uni-bonn.de}
\author{S.~Grab}
\email[]{sgrab@th.physik.uni-bonn.de}
\affiliation{Physikalisches Institut, University of Bonn, Bonn, Germany}

\author{P.~Richardson}
\email[]{Peter.Richardson@durham.ac.uk}

\affiliation{IPPP, University of Durham, Durham, UK}
\affiliation{Theoretical Physics Group, CERN, Geneva, Switzerland}

\begin{abstract}
We calculate the total cross section for single charged slepton
production in association with a top quark at hadron colliders in the
baryon triality (B$_3$) supersymmetric model. We compute event rates
for the Tevatron and LHC. We study the signatures for
different supersymmetric scenarios including neutralino and stau
LSPs. We perform a detailed analysis with basic
cuts for the B$_3$ operator $\lam^\prime_{231}$ using Monte Carlo simulations
to show that the signal can be distinguished from the background at the LHC. In
particular we employ the resulting lepton charge asymmetry.
\end{abstract}

\preprint{BONN-TH-2007-08} 
\preprint{CERN-PH-TH-2007-028}
\preprint{DCPT/08/16} 
\preprint{IPPP/08/08}

\maketitle
\input{axodraw.sty}

\section{Introduction}

Supersymmetry (SUSY) \cite{Wess:1974tw} is a widely considered
extension of the Standard Model~(SM) of particle physics
\cite{SM}. If it exists, it is necessarily broken, with a mass scale
of order of the TeV energy scale \cite{hierarchy}. This energy region is probed
at both the Tevatron and LHC; the search for
SUSY is therefore of paramount interest \cite{susy-lhc,drees}. It is the
purpose of this paper to consider a specific supersymmetric production
mechanism and investigate its viability at the Tevatron and LHC.

\mymed

The general renormalisable superpotential with minimal particle
content \cite{ssm-superpot} includes the following lepton or baryon
number violating interactions,
\bea
W_{\not\mathrm{P}_6} & = & \eps_{ab}\left[\frac{1}{2}\lam_{ijk}  L_i^aL_j^b
\ovl{E}_k + \lam'_{ijk}L_i^aQ_j^{bx}\ovl{D}_{kx}\right]\notag \\
&&\hspace{-0.2cm}
+\eps_{ab}\kap^i  L_i^aH_2^b
+\frac{1}{2}\eps_{xyz}{\lam}''_{ijk}
\ovl{U}_i^{\,x} \ovl{D}_j^{\,y} \ovl{D}_k^{\,z} \,,
\label{P6v-superpot}
\eea
where we have employed the standard notation of
Ref.~\cite{Dreiner:1997uz} for the superfields, couplings and 
indices. If all terms are simultaneously present, they lead to rapid
proton decay \cite{proton-decay}.  Therefore SUSY, must be augmented
by an additional symmetry. The discrete anomaly-free gauge symmetries
R-parity \cite{Farrar:1978xj} and proton hexality, $\Psix$
\cite{Dreiner:2005rd}, forbid all of the above terms. However,
R-parity does not forbid the dangerous dimension-five proton decay
operators \cite{ssm-superpot}.

\mymed

An equally well motivated solution to the proton decay problem is
baryon triality, B$_3$, a discrete anomaly-free $\boldsymbol{Z}_3
$-symmetry, which prohibits the $\bar{U}\bar{D}\bar{D}$ operator in
Eq.~(\ref{P6v-superpot}) \cite{Ibanez:1991pr,Ibanez:1991hv,
Dreiner:2006xw}. This solution has an additional feature, it naturally
leads to small neutrino masses \cite{Hall:1983id,Hempfling:1995wj,
Borzumati:1996hd,Hirsch:2000ef,Allanach:2003eb,Dreiner:2007uj,
Allanach:2007qc}, as experimentally observed \cite{neutrino-exp}.
Furthermore, B$_3$ supersymmetric models which include also a 
candidate for dark matter can for example be realized within the UMSSM
\cite{UMSSM}.

\mymed

The baryon-triality collider phenomenology has three main
distinguishing features, compared to the $\Psix$ case
\cite{Dreiner:1997uz,Barbier:2004ez}:
\begin{enumerate}
\item the lightest supersymmetric particle (LSP) is not stable and 
can decay in the detector. It also need not be the lightest 
neutralino;
\item SUSY particles are also produced singly, possibly on resonance; 
\item lepton flavour and number are violated.
\end{enumerate}
These lead to dramatically different signatures at hadron colliders
\cite{Dreiner:1991pe,Barbier:2004ez,Allanach:1999bf,Allanach:2006st} 
compared to the more widely studied P$_6$ case.

\mymed

In the following, we focus solely on the signatures due to a
non-vanishing $L_iQ_j\bar D_k$ operator. At hadron colliders
this allows resonant single slepton and sneutrino production via
incoming quarks,
\bea
\bar u_j + d_k & \overset{\lamp_{ijk}}{\longrightarrow} & \tilde\ell_i^-,
\label{eq:singleslep} \\
\bar d_j + d_k & \overset{\lamp_{ijk}}{\longrightarrow} & \tilde\nu_i
\;. \label{sneutrino_production}
\eea
Here: $u_j$ and $d_k$ denote up and down type quarks of generations
$j$ and $k$, respectively; a bar denotes an anti-quark; $\tilde\ell_i
^-$ and $\tilde\nu_i$ denote negatively charged sleptons and
sneutrinos of generation $i$, respectively. 

\mymed

The tree-level processes Eqs.~(\ref{eq:singleslep}),
(\ref{sneutrino_production}) were first considered in
Refs.~\cite{Dimopoulos:1988jw,Dimopoulos:1988fr}. Like-sign dilepton
events or three lepton final states were considered in
Refs.~\cite{Dreiner:2000vf,Moreau:2000bs,Deliot:2000mf}. These papers
assume a neutralino LSP, which can decay leptonically via $\lam'_
{ijk}$, e.g. $\tilde{\chi}_1^0 \rightarrow \ell_i^+ \bar u_jd_k$. The
case of a gravitino LSP was considered in
Ref.~\cite{Allanach:2003wz}. The process has also been studied by the
D0 collaboration at the Tevatron \cite{Abazov:2002es,Abazov:2006ii},
for the operator $L_2 Q_1 \bar D_1$ and a neutralino LSP, setting
limits on the relevant masses and couplings. In
Refs.~\cite{Choudhury:2002au,Yang:2005ts,Dreiner:2006sv,Chen:2006ep}
the NLO-QCD corrections were considered and in Ref.
\cite{Dreiner:2006sv} the SUSY-QCD corrections were taken into 
account. Gluon fusion contributions were included in
Ref.~\cite{Chen:2006ep}.

\mymed

The case $j=3$ in Eq.~(\ref{eq:singleslep}) is special, as there are
no top quarks in the incoming proton. Instead, one must consider the
production of a single slepton in association with a SM particle.
Several mechanisms for associated single supersymmetry production,
\textit{e.g.} $d_j\bar d_k\ra\tilde{\chi}^+_1\ell^-_i$, have been
studied in the literature, see for example
Refs.~\cite{Borzumati:1999th,Moreau:1999bt,Deliot:2000mf,
Chaichian:2003kf,Allanach:1997sa}. In the following, we investigate in
detail the case of the operator $L_iQ_3\bar D_k$. Here, single charged
slepton production is only possible in association with a top
quark. Before studying the phenomenological details, we first recall
the strongest experimental bounds on the couplings $\lam'_{i3k}$ at
the $2\sigma$ level. They are shown in Table~\ref{lamp_bounds}
\cite{Dreiner:1997uz,Allanach:1999ic,Barbier:2004ez,Chemtob:2004xr,Dreiner:2006gu}.
\begin{table}[tb]
\begin{ruledtabular}
\begin{tabular}{c|c||c|c}
$\lam'_{131}$ & $0.019 \times (m_{\tilde{t}_L}/100\gev)$  & 
$\lam'_{132}$ & $0.28 \times (m_{\tilde{t}_L}/100 \gev)$  \\
$\lam'_{231}$ & $0.18 \times (m_{\tilde{b}_L}/100 \gev)$  &
$\lam'_{232}$ & $0.45 \, (m_{\tilde{s}_R}=100 \, \text{GeV}) $\\
$\lam'_{331}$ & $0.45 \,\, (m_{\tilde{q}}=100 \, \text{GeV})$  & 
$\lam'_{332}$ & $0.45 \,\, (m_{\tilde{q}}=100 \, \text{GeV})$  \\
$\lam'_{i33}$ & $\mathcal{O}(10^{-4})$  &    
\end{tabular}
\caption{\label{lamp_bounds}Upper $2\sigma$ bounds on $\lam'_{i3k}$. 
The strong bounds on $\lam'_{i33}$ stem from neutrino masses $m_{\nu}$, 
assuming $m_{\nu}<1\,\text{eV}$ and left right mixing in the sbottom sector.
The limits depend on the squark masses, $m_{\tilde{q}_{L(R)}}$ is
the mass of the left (right) handed squark $\tilde{q}_{L(R)}$.}
\end{ruledtabular}
\end{table} 
We neglect bounds, which assume a specific
(Standard Model) quark mixing between the three generations
\cite{Agashe:1995qm} or bounds using the renormalization group running
of $\lam'_{i3k}$ \cite{Allanach:1999ic,Allanach:1999mh,Allanach:2003eb}.  

\mymed

\begin{figure}[t!] 
  \setlength{\unitlength}{1cm} \includegraphics[scale=0.5]{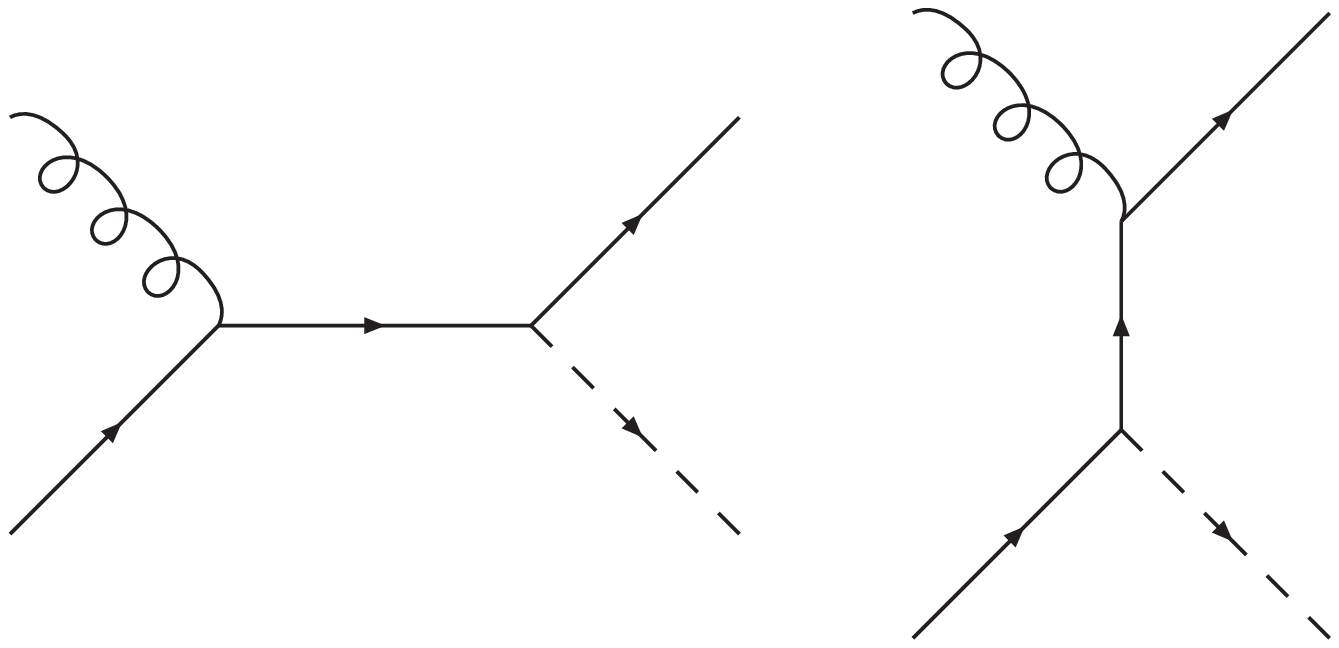}
  \put(-6.5,2.9){$g$}
  \put(-6.5,0.5){$d_k$}
  \put(-3.6,2.9){$t$}
  \put(-3.6,0.5){$\tl_i^-$}
  \put(-5.0,1.9){$d_k$}
  \put(-2.2,2.5){$g$}
  \put(-2.2,0.9){$d_k$}
  \put(-0.3,2.5){$t$}
  \put(-0.3,0.9){$\tl_i^-$}
  \put(-0.9,1.6){$t$}
\caption{\label{topstau} Feynman diagrams contributing to the partonic 
process $g+d_k\ra t+\tl_i^-$.}
\end{figure}

At leading order there are two production mechanisms for 
slepton production in association with a top quark. First the
Compton-like processes
\begin{subequations}
\label{compton_process}
\begin{eqnarray}
g + d_k &\ra& \tilde{\ell}_{i}^- + t\,,  \label{compton_process1}\\ 
g + \bar d_k &\ra& \tilde{\ell}_{i}^+ + \bar t\,.   
\label{compton_process2}       
\end{eqnarray}
\end{subequations}
The relevant leading-order diagrams are given in Fig.~\ref{topstau}.
Here, $g$ denotes an incoming gluon in the proton and $t$ a final-state
top quark. 

\mymed

The second slepton production mechanism is $t \overline{t}$ pair
production followed by the $t$ or $\bar{t}$ decaying into $\tl_i^+$
or $\tl_i^-,$ respectively. The main production mechanisms for
$t\bar t$ production, at $O(\alpha_s^2)$, are
\begin{equation}
\left.\begin{array}{rcl}
q + \bar q &\ra& t + \bar t  \\ 
g + g &\ra& t + \bar t
\end{array}\;\right\}\,,\qquad t \ra \tilde\ell_i^+ + d_k
\label{ttbar_process}
\end{equation}
where $q$ ($\bar q$) is a (anti-)quark. This is only kinematically
allowed if
\begin{equation}
m_t > m_{\tilde\ell i} + m_{d_k}\,.
\end{equation}
Since, as we shall see, the branching fraction for the B$_3$ top quark
decay is small, we only consider one B$_3$ decay, for either the top
or the anti-top quark.

\mymed

In Ref.~\cite{Borzumati:1999th}, single slepton production was
considered for the specific case of $\lam'_{333}\not = 0$. This
process is however disfavoured due to the strict bound on the relevant
coupling from neutrino masses, \textit{cf}. 
Table~\ref{lamp_bounds} \footnote{Even if one assumes that the
one-loop contributions to the neutrino mass are suppressed, two loop
contributions lead to a bound on $\lam'_{333}$ of $10^{-2}$
\cite{Accomando:2006ga, Borzumati:2002bf}, which still leads to a
suppressed cross section.}. Thus the work was extended to the
couplings $\lam'_{331}$ and $\lam'_{332}$ \cite{Borzumati:2002bf}. We
go beyond this work to include a signal over background analysis. We
also present the analytic formula for the cross section,
Eq.~(\ref{partonic_xsection}), for the first time, and analyse the
resulting signatures. We give a detailed phenomenological analysis for
the special case $\lam^\prime_{231}$ which can be generalized to
$\lam^\prime_{131}$.

\mymed

In Ref.~\cite{Belyaev:2004qp}, top quark pair production and
subsequent top decay via $\lam'_{i3k}$ was considered. Off-shell
top quark effects were also taken into account. A signal over
background analysis was performed for two scenarios. The first
scenario assumed maximal stop-scharm mixing. It was pointed out that
associated slepton production with slepton masses 150 GeV and 200 GeV
can be measured, depending on the magnitude of $\lam'_{i3k}$. The
second scenario assumed no flavour violation in the squark
sector. Ref.~\cite{Belyaev:2004qp} claimed that in this regime
sleptons with mass 200 GeV can not be measured. We go beyond the work
of \cite{Belyaev:2004qp}. We show that it is possible to detect 
associated slepton production even for slepton masses larger than 300
GeV, if $\lam^\prime_ {231}$ or $\lam^\prime_{131}$ is of $\mathcal{O}
(0.1)$. We will achieve this with the help of the Compton-like process 
(\ref{compton_process}).

\mymed
 
The outline of this paper is as follows. In Sect.~\ref{production} we
calculate the cross section for the production of a charged slepton in
association with a top quark, at leading order. In Sect.~\ref{events}
we systematically present the possible resulting signatures at the
LHC. In Sect.~\ref{details} we discuss in detail a case study for the
operator $\lam_{231}^\prime L_2Q_3\bar D_1$.  We study the dominant
$t\bar{t}$ and $W^\pm$ backgrounds. Using the HERWIG Monte
Carlo program \cite{Corcella:2000bw,Corcella:2002jc,Dreiner:1999qz},
we devise a set of cuts in order to distinguish the two. We do not
include a simulation of the detector. Our conclusions are presented in 
Sect.~\ref{conclusion}.


\section{Single slepton production via $\boldsymbol{L_i Q_3 \bar D_k}$}
\label{production}

\subsection{Partonic cross sections}

The spin and colour averaged matrix element squared for the
Compton-like process Eq.~(\ref{compton_process}) is given at leading order by
\begin{align}
&|\overline{M}|^2= \frac{\pi\lam'^2_{i3k} \alpha_s C_F |L^{\ell_i}_
{1\alpha}|^2}{4} \Bigg\{
\frac{m_t^2-\hat{t}}{\hat{s}} \nonum \\  
&+\frac{(m^2_{\tl i}-m_t^2)(m^2_{\tl i}-\hat{s}-\hat{t})-
(3m_t^2-m^2_{\tl i}+\hat{s})(\hat{t}-m^2_{\tl i})}{(\hat{t}-m^2_{t})^2} \nonum \\
&\quad+ \frac{2[m_t^2 \hat{s}+(\hat{t}-m^2_{\tl i}) (m^2_{\tl i}-m_t^2-\hat{s})]}
{\hat{s}(\hat{t}-m_t^2)} \Bigg\},
\end{align}
where $\alpha_s$ is the QCD coupling constant, $C_F=4/3$ is the quadratic
Casimir of $SU(3)_c$, $m_{\tl i}$ is the mass of the slepton
and $L_{1\alpha}^{\ell_i}$ is the relevant matrix
element of the left-right slepton mixing matrix. The explicit form as
a function of the mixing angle is given, for example, in
Ref.~\cite{Dreiner:1999qz}. In
accordance with the parton model, we have neglected the mass of $d_k$.
We have made use of the partonic Mandelstam variables
\begin{subequations}
\begin{align}
\hat{s} &= (d_k+g)^2 = (t+\tl_i)^2\,, \\
\hat{t} &= (d_k-\tl_i )^2 = (g - t)^2\,,
\end{align}
\end{subequations}
where we denote the particle four momenta by the particle letter.
Integrating over phase space, we obtain the total partonic cross 
section:
\begin{align}
\hat{\sigma}=&\frac{\lam'^2_{i3k} \alpha_s C_F |L^{\ell_i}_{1\alpha}|^2}
{64 \hat{s}^2} \Bigg\{
\frac{1}{2\hat{s}} \left[ 2m_t^2(\hat{t}_+ - \hat{t}_-)-(\hat{t}_+^2 - \hat{t}_-^2) \right] \nonumber \\
&+(\hat{s} + 2 m_t^2) \ln \left( \frac{\rho_-}{\rho_+} \right) 
+\frac{2 m_t^2(m_{\tl i}^2-m_t^2)(\hat{t}_+-\hat{t}_-)}{\rho_+\rho_-} \nonumber \\
&+\frac{2(m_{\tl i}^4+m_t^4-2m_t^2m_{\tl i}^2-m_{\tl i}^2\hat{s})}{\hat{s}} \ln \left( \frac{\rho_-}{\rho_+} \right) \nonumber \\
&+\frac{2(\hat{t}_+-\hat{t}_-)(m_{\tl i}^2-m_t^2-\hat{s})}{\hat{s}} \Bigg\},
\label{partonic_xsection}
\end{align}
where
\begin{align}
\rho_{\pm}&=m_t^2-\hat{t}_{\pm}, \\
\hat{t}_{\pm}&= m^2_{\tl i}-\frac{1}{2} [\hat{s}+m^2_{\tl i}-m_t^2 
\mp \lam^{\frac{1}{2}}(\hat{s},m^2_{\tl i},m_t^2)],
\end{align}
with the phase-space function given by $\lam(x,y,z)=x^2+y^2+z^2-2xy-
2xz-2yz$.

\mymed

The tree-level partonic matrix element squared for top quark pair
production is given for example in Ref.~\cite{Combridge:1978kx}. We
shall only consider on-shell top quark pair production. The slepton
then arises through the decay of a real top quark. In order to obtain
the signal rate, we thus also require the partial decay width of the
top quark, via the $L_iQ_3\bar D_k$ operator. It is given by
\begin{align}
\Gamma_{t\ra d_k \tl_i^+} =& \frac{\lam'^2_{i3k} 
|L^{\ell_i}_{1\alpha}|^2 }{32 \pi m_t}
\left(1+\frac{m_{d_k}^2}{m_t^2}-\frac{m_{\tl i}^2}{m_t^2} \right) \nonumber \\
& \times \lam^{1/2}(m_t^2,m_{d_k}^2,m_{\tl i}^2)\;.
\label{partial_width}
\end{align} 
See also Refs.~\cite{Borzumati:1999th,Dreiner:1991dt,top_decay,
Agashe:1995qm,Dreiner:1999qz}.  We obtain a branching ratio of
$8.2\times10^{-4}$ for the R-Parity violating top decay
(\ref{partial_width}) for $\lam'_{i3k}=0.1$, $m_t=175 \, \text{GeV}$,
top width $\Gamma_t=1.5 \, \text{GeV}$ and $m_{\tilde{\ell} i}=150\,
\text{GeV}$. We neglect the mass of $d_k$ and set
$L^{\ell_i}_{1\alpha}=1$.

\subsection{Total Hadronic cross section}

\begin{figure}[h!] 
  \setlength{\unitlength}{1cm} \includegraphics[scale=0.55, bb =  40 70
  450 530, clip=true]{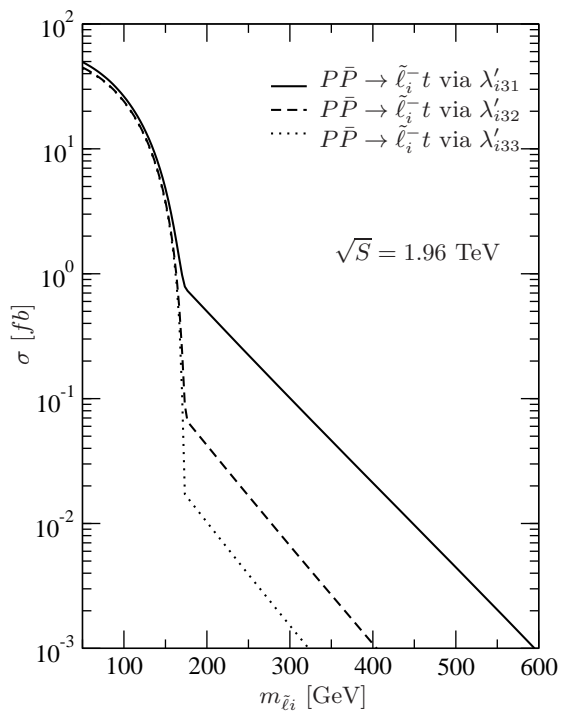}
	\put(-3.6,5.6){$\sqrt{S}=1.96$ TeV}
	\put(-3.8,7.9){$P \bar P \ra \tilde{\ell}_{i}^- t$ via $\lam'_{i31}$}
	\put(-3.8,7.5){$P \bar P \ra \tilde{\ell}_{i}^- t$ via $\lam'_{i32}$}
	\put(-3.8,7.1){$P \bar P \ra \tilde{\ell}_{i}^- t$ via $\lam'_{i33}$}
	\put(-7.9,4.2){\rotatebox{90}{$\sigma$ [$fb$]}}
	\put(-4.6,-0.3){$m_{\tl i}$ [GeV]}
\caption{\label{slepitop_xsection_Tevatron} Single slepton production in
  association with a top at the Tevatron. The cross sections 
for $\tilde{\ell}_{i}^+ \bar t$ production are equal to the cross sections 
for $\tilde{\ell}_{i}^- t$ production.}
\end{figure}

\begin{figure}[h!] 
  \setlength{\unitlength}{1cm} 
\includegraphics[scale=0.55, bb = 40 70
  450 530, clip=true]{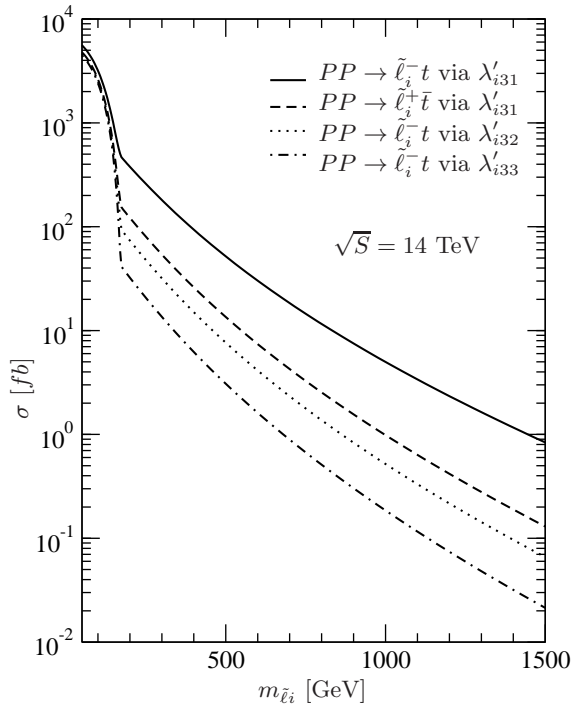}
  \put(-3.6,5.6){$\sqrt{S}=14$ TeV} \put(-3.8,7.9){$P P \ra
  \tilde{\ell}_{i}^- t$ via $\lam'_{i31}$} \put(-3.8,7.5){$P P
  \ra \tilde{\ell}_{i}^+ \bar t$ via $\lam'_{i31}$}
  \put(-3.8,7.1){$P P \ra \tilde{\ell}_{i}^- t$ via
  $\lam'_{i32}$} \put(-3.8,6.7){$P P \ra \tilde{\ell}_{i}^-
  t$ via $\lam'_{i33}$} \put(-7.9,3.4){\rotatebox{90}{$\sigma$
  [$fb$]}} \put(-4.6,-0.3){$m_{\tl i}$ [GeV]}
  \caption{\label{slepitop_xsection_LHC} Same as 
Fig.~\ref{slepitop_xsection_Tevatron}, but for the LHC. The 
  cross section for $\tilde{\ell}_{i}^+\bar t$ production via $\lam
  '_{i32}$ ($\lam'_{i33}$) is equal to the cross section for $\tilde
  {\ell}_{i}^- t$ production via $\lam'_{i32}$ ($\lam'_{i33}$), as 
  it always involves incoming sea quarks.}
\end{figure}

In Fig.~\ref{slepitop_xsection_Tevatron}
(Fig.~\ref{slepitop_xsection_LHC}), we show the hadron level cross
section at the Tevatron (LHC) for single slepton production in
association with a top quark, as a function of the slepton mass including
both production mechanisms. We set $\lam'_{i3k}=0.1$ and assume it is the
only non-vanishing B$_3$ coupling. We vary the index $k$ and the
charge of the final state slepton, which correspond to different
parton density functions (PDFs). Here we use the CTEQ6L1 PDFs
\cite{Pumplin:2005rh}. The renormalisation, $\mu_R$, and 
factorisation, $\mu_F$, scales are taken to be equal,
$\mu_R=\mu_F= m$, where $m\equiv 2m_t$ [$\equiv m_{\tl i}+m_{t}$]
in the case of slepton production via a $t \bar t$
pair (\ref{ttbar_process}) [via the Compton-like process
(\ref{compton_process})].  Furthermore, we set the left-right slepton 
mixing matrix element $L^{\ell_i}_{1\alpha}$ equal to one.  Results
for other values of $\lam'_{i3k}$ and mixing matrix elements
$L^{\ell_i}_{1\alpha}$ are easily obtained by rescaling according to
Eqs.~(\ref{partonic_xsection}) and (\ref{partial_width}).  The top
mass is taken to be 175 GeV and the total (SM) top quark decay width
to be 1.5 GeV.  We take $m_{d_3}=m_b=4.5 \,\text{GeV}$, if we have a
b quark in the final state and neglect the masses of the d and
s quarks.

\mymed

In both figures, we see a kink in the cross section when $m_{\tl i}=
m_t - m_{d_k}$. For smaller slepton masses the top quark pair
production mechanism dominates; for larger masses the Compton-like
processes dominate, since the slepton can no longer be produced
on-shell in top decay.

\mymed

\begin{figure}[ht!] 
  \setlength{\unitlength}{1cm} \includegraphics[scale=0.55, bb =  40 70
  450 530, clip=true]{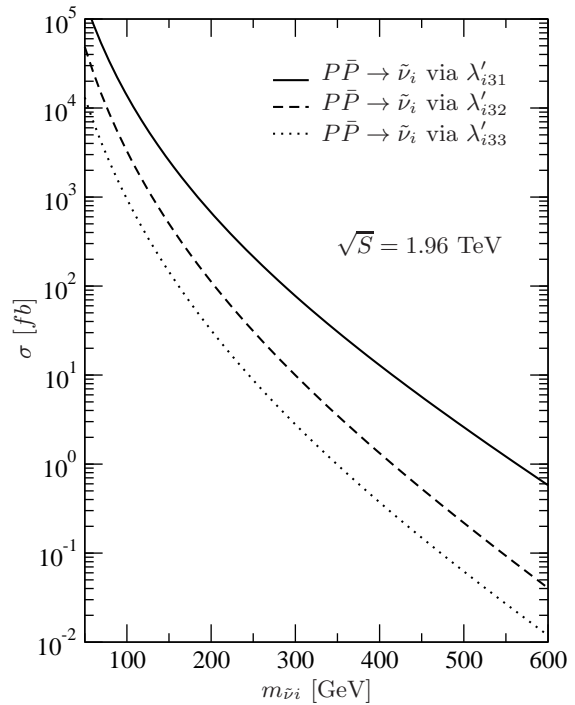}
	\put(-3.6,5.6){$\sqrt{S}=1.96$ TeV}
	\put(-3.8,7.9){$P \bar P \ra \tilde{\nu}_{i}$ via $\lam'_{i31}$}
	\put(-3.8,7.5){$P \bar P \ra \tilde{\nu}_{i}$ via $\lam'_{i32}$}
	\put(-3.8,7.1){$P \bar P \ra \tilde{\nu}_{i}$ via $\lam'_{i33}$}
	\put(-7.9,4.2){\rotatebox{90}{$\sigma$ [$fb$]}}
	\put(-4.6,-0.3){$m_{\tilde{\nu} i}$ [GeV]}
\caption{\label{xsection_sneutrino_Tevatron} Single sneutrino production
cross section at the Tevatron. 
The cross sections for $\tilde{\nu}_{i}^*$ production 
are equal to the cross sections for $\tilde{\nu}_{i}$ production.}
\end{figure}

\begin{figure}[ht!] 
  \setlength{\unitlength}{1cm} \includegraphics[scale=0.55, bb =  40 70
  450 530, clip=true]{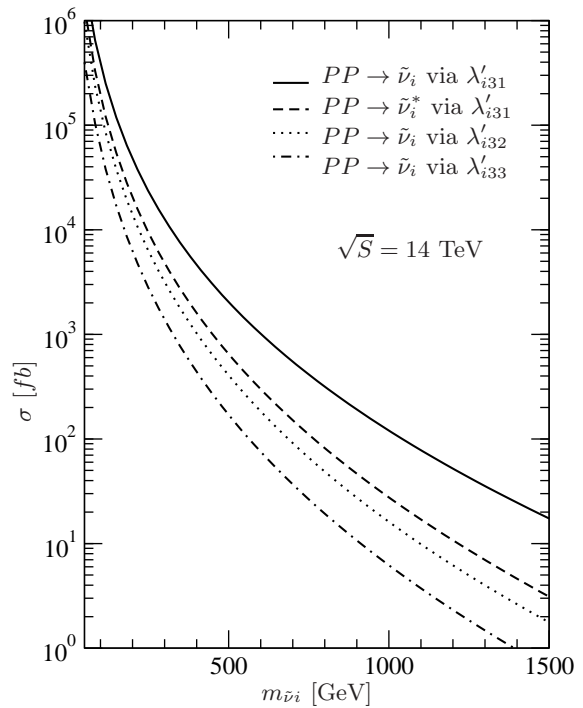}
	\put(-3.6,5.6){$\sqrt{S}=14$ TeV}
	\put(-3.8,7.9){$P P \ra \tilde{\nu}_{i}$ via $\lam'_{i31}$}
	\put(-3.8,7.5){$P P \ra \tilde{\nu}_{i}^*$ via $\lam'_{i31}$}
	\put(-3.8,7.1){$P P \ra \tilde{\nu}_{i}$ via $\lam'_{i32}$}
	\put(-3.8,6.7){$P P \ra \tilde{\nu}_{i}$ via $\lam'_{i33}$}
	\put(-7.9,3.4){\rotatebox{90}{$\sigma$ [$fb$]}}
	\put(-4.6,-0.3){$m_{\tilde{\nu} i}$ [GeV]}
\caption{\label{xsection_sneutrino_LHC} Same as Fig. 
\ref{xsection_sneutrino_Tevatron}, but for the LHC.
The cross section for $\tilde{\nu}_{i}^*$ production
via $\lam'_{i32}$ ($\lam'_{i33}$) is equal to the cross section 
for $\tilde{\nu}_{i}$ production via $\lam'_{i32}$ ($\lam'_{i33}$),
since only initial-state sea quarks are involved.}
\end{figure}

\begin{figure}[ht!] 
  \setlength{\unitlength}{1cm} \includegraphics[scale=0.4, bb = 65 93
  580 530, clip=true]{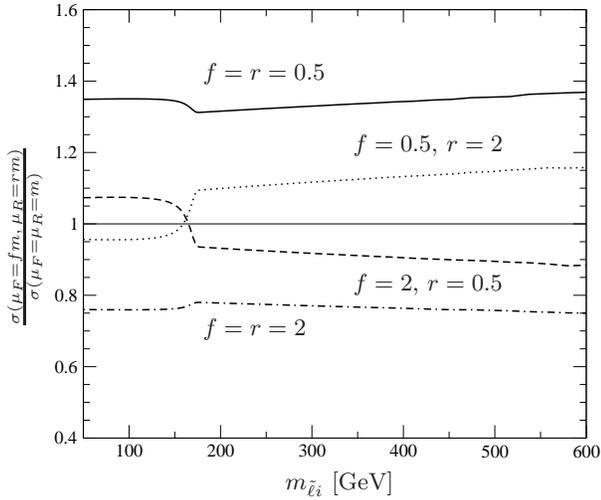}
	\put(-5.3,5.1){$f=r=0.5$}
	\put(-3.3,4.15){$f=0.5,\,r=2$}
	\put(-3.3,2.3){$f=2,\,r=0.5$}
	\put(-5.3,1.7){$f=r=2$}
  	\put(-7.9,1.8){\rotatebox{90}
	{$\frac{\sigma(\mu_F=fm,\,\mu_R=rm)}{\sigma(\mu_F=\mu_R=m)}$}}
	\put(-4.2,-0.35){$m_{\tl i}$ [GeV]}
\caption{\label{scale_Tevatron} Factorisation scale $\mu_F=f\cdot m$ and 
renormalisation scale $\mu_R=r\cdot m$ dependence of the hadronic
$\tilde{\ell}_{i}^- t$ production cross section via $\lam'_{i31}$ 
at the Tevatron. $\mu_F$ and $\mu_R$ are independently
taken equal to 2 and 0.5 times $m$, where $m\equiv 2m_t$ ($\equiv
m_{\tl i}+m_{t}$) in the case of slepton production via a $t \bar t$
pair \eqref{ttbar_process} (via the Compton-like process 
\eqref{compton_process}~).}
\end{figure}

\begin{figure}[ht!] 
  \setlength{\unitlength}{1cm} \includegraphics[scale=0.4, bb = 65 93
  590 540, clip=true]{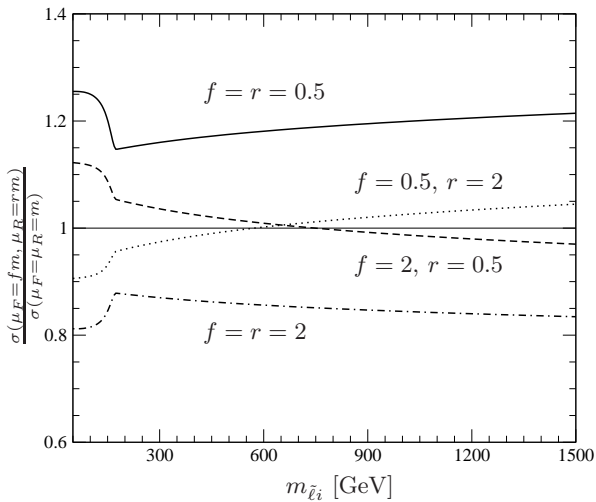}
	\put(-5.3,4.9){$f=r=0.5$}
	\put(-3.3,3.7){$f=0.5,\,r=2$}
	\put(-3.3,2.6){$f=2,\,r=0.5$}
	\put(-5.3,1.7){$f=r=2$}
  	\put(-7.9,1.6){\rotatebox{90}
	{$\frac{\sigma(\mu_F=fm,\,\mu_R=rm)}{\sigma(\mu_F=\mu_R=m)}$}}
	\put(-4.2,-0.35){$m_{\tl i}$ [GeV]}
        \caption{\label{scale_LHC} Same as Fig. \ref{scale_Tevatron}
		but for the LHC.}
\end{figure}

For comparative discussions later,
Fig.~\ref{xsection_sneutrino_Tevatron}
(Fig.~\ref{xsection_sneutrino_LHC}) shows the NLO hadronic cross
section for resonant sneutrino production, \textit{cf}.
Eq.~(\ref{sneutrino_production}), at the Tevatron (LHC) via
$\lam'_{i3k}=0.1$, including NLO QCD corrections \cite{foot1}. We
employ the $\overline{\text{MS}}$ renormalisation scheme and the (NLO)
CTEQ6M PDFs \cite{Pumplin:2005rh}. The renormalisation and factorisation
scales are taken to be the sneutrino mass, $\mu_R=\mu_F=
m_{\tilde{\nu}_i}$.

\mymed

In Fig.~\ref{slepitop_xsection_Tevatron}, we see that at the Tevatron,
even for small slepton masses, $m_{\tl i}= 100\,\mathrm{GeV}$, we
expect only 25 (25) charged slepton events with negative (positive)
charge, \textit{i.e.} $\tl_i^-$ ($\tl_i^+$), for an integrated
luminosity of $1\,\mathrm{fb}^{-1}$ and the (relatively large)
coupling $\lam'_{i31}=0.1$.  The cross section is dominated by the
$t\bar t$ pair production (\ref{ttbar_process}). Only $10\%$ of the
above sleptons at the Tevatron are produced by the Compton-like
process (\ref{compton_process}).  At the Tevatron, the cross section
is symmetric in the slepton charge due to the charge symmetry of the
incoming state.

\mymed

As we can see in Fig.~\ref{slepitop_xsection_LHC}, we have a
significantly larger hadronic cross section at the LHC for a given
slepton mass. In particular, for $m_{\tl i}=100\,\mathrm{GeV}$ and $
\lam'_{i31}=0.1$ the LHC will produce more than 31 000 (26 000)
sleptons $\tl_i^-$ ($\tl_i^+$) for an integrated luminosity of $10\,
\mathrm{fb}^{-1}$\!. $27 \%$ ($11\%$) of these sleptons are
produced via the Compton-like process. For the same coupling and for
$m_{\tilde{\nu}_i}= 100 \,\text{GeV}$, we will produce approximately
14~000 sneutrinos at the Tevatron
(Fig.~\ref{xsection_sneutrino_Tevatron}) for $1 \, \text{fb}^{-1}$ and
3~800~000 at the LHC (Fig.~\ref{xsection_sneutrino_LHC}) for $10 \,
\text{fb}^{-1}$, via the partonic process
Eq.~(\ref{sneutrino_production}). Thus, depending on the decays, we
might expect this to be the discovery mode, for equal supersymmetric
masses. Here we focus on the potential of the charged slepton
production cross section.

\mymed

For heavier charged sleptons, $m_{\tl i}=800\gev$, we expect no events
at the Tevatron and more than 110 (25) $\tl_i^-$ ($\tl_i^+$) events at
the LHC with $10\,\mathrm{fb}^{-1}$. Above the threshold of $ m_{\tl
i}=m_t-m_{d_k}$, practically all slepton events are produced via the
Compton-like process, since the other process only proceeds via
off-shell top quarks. The cross section is so small because the parton
luminosity is too small at the required high values of the
proton/anti-proton fractional momenta, $x\gsim 0.1$. This situation
changes at the LHC, where we probe significantly smaller values, $x<
0.1$, for the same slepton mass. Furthermore, the Tevatron will
produce no sneutrinos, for $\lam'_{i31}= 0.1$, and $m_{\tilde{\nu}
i}=800 \,\text{GeV}$.  For the same set of B$_3$ parameters, the LHC
will produce about 3~200 sneutrinos for $10 \, \text{fb}^{-1}$.

\mymed

At the LHC, there is an asymmetry between the hadronic cross sections
for $\tl_i^-$ and $\tl_i^+$ production via the $L_i Q_3\bar D_1$
operator ($k=1$!). This is perhaps not surprising, as the initial
state is asymmetric under charge reversal. In the case of the
Compton-like process (\ref{compton_process}), the asymmetry is due to
the negatively charged slepton being produced by an incoming valence
$d$-quark, while the positively charged slepton is produced by a $\bar
d$ sea quark. The latter has a lower luminosity in the proton. In
Sect.~\ref{details} we will use this asymmetry to separate the B$_3$
process from the SM background.

\mymed

In order to estimate the influence of higher order corrections on the
production cross section, we vary the renormalisation and
factorisation scales independently between $m/2$ and $2m$. At the
Tevatron, Fig.~\ref{scale_Tevatron}, the hadronic cross section for
$\tilde{\ell}_{i}^- t$ production via $\lam'_{i31}$ changes by up to
$40\%$.  At the LHC, Fig.~\ref{scale_LHC}, the scale uncertainties are
reduced to approximately $25\%$. In the domain where $m_{\tl
i}<m_t-m_{d_k}$, we have a stronger dependence on the renormalisation
scale compared to $m_{\tl i}>m_t-m_{d_k}$, because $t\bar t$
production is ${\cal O}(\alpha_s^2(\mu_r))$.  According to
Ref.~\cite{Bonciani:1998vc,Cacciari:2003fi}, NLO-QCD corrections,
including a NLL resummation, increases the $t\bar t$ production cross
section by approximately $40\%$ ($80\%$) at the Tevatron (LHC).

\mymed

The hadronic cross section for single stau, $\tilde{\tau}$, production
via a non-vanishing $\lam'_{333}$ coupling, was also considered by
Borzumati {\it et. al.} \cite{Borzumati:1999th}. There, the $2\ra2$
processes, Eqs.~(\ref{compton_process}) were included, together with
the (tree-level) $2\ra3$ slepton-strahlung processes
\begin{equation}
\label{two_to_three}
\left.\begin{array}{c}
g+g \vspace{0.15cm} \\
q+\bar q 
\end{array}\right\}
                     \ra t+\bar b+\tilde{\tau}^-\, ,
\end{equation}
which they show, for $m_{\tilde{\tau}}<m_t-m_{b}$, to be equivalent to
the $2\ra2$ processes, Eqs.~(\ref{ttbar_process}). The $\bar b$ and
$\tilde{\tau}^-$ are produced via a virtual top.  They employed the
CTEQ4L \cite{Lai:1996mg} PDFs and all matrix elements were multiplied
by the CKM factor $V_{tb}$.  We have calculated the hadronic cross
sections using the same PDFs and the same parameter set
\cite{private_communication}. We co\"incide exactly, where single
slepton production is dominated by the $t\bar t$ process, \textit{i.e}
for $m_{\tilde{\tau}}<m_t-m_{b}$. For $m_{\tilde{\tau}}>m_t-m_{b}
$, we underestimate the total cross section at the Tevatron by $20 \%$
for $m_{\tilde{\tau}}=300\,\mathrm{GeV}$ and by a factor of roughly
two for $m_{\tilde{\tau}}=200$ GeV, compared to
Ref.~\cite{Borzumati:1999th}.  In this region the above $2
\rightarrow 3$ processes, where the slepton is produced by a
quark-antiquark pair, can give the main contribution compared to the
$gb \rightarrow\tilde{\tau} t$ partonic process, where a gluon and/or
sea-quark is needed with large Bjorken x. However, in this region where
there are large discrepancies practically no sleptons are produced at
the Tevatron. Our prediction for the LHC differs by $+30\%$ for $
m_{\tilde{\tau}} > m_t- m_{b}$. 

\mymed

Borzumati {\it et. al.} extended their analysis to the $\lam'_{332}$
and $\lam'_{331}$ couplings \cite{Accomando:2006ga}. They present the
results for the $2 \rightarrow 2$ process (\ref{compton_process}) and
the $2\rightarrow 3$ process (\ref{two_to_three}), separately.  For
$m_{\tilde{\tau}} < m_t - m_{d_k}$, we co\"incide exactly at the
Tevatron as well as at the LHC. For $m_{\tilde{\tau}} > m_t -
m_{d_k}$, our predictions co\"incide exactly with their cross section
predictions for the $2 \rightarrow 2$ process. Furthermore, it is
shown in \cite{Accomando:2006ga} that for $m_{\tilde{\tau}} > m_t -
m_{d_k}$, the $2 \rightarrow 3$ contributions are small or even
negligible. At the Tevatron, the $2\rightarrow 3$ process contributes
roughly $35\%$ ($5\%$) to the total hadronic cross section for
$\lam'_{332}\not =0$ ($\lam'_{331}\not =0$).  At the LHC these
contributions are $25\%$ ($5\%$). The reason is that the
cross sections induced by the $2\ra3$ process have similar sizes for
any value of $k$. But the $2\ra2$ process for $\lam'_{332}\not =0$
($\lam'_{331}\not =0$) is enhanced by a factor of $5$ ($\gsim 10$) due
to a s-quark (valence d-quark) in the initial state.

\mymed

We conclude, that our LO approximation is valid in the
phenomenologically relevant region, where one is able to produce a
single slepton in association with a top quark. 
We have not included the $2\ra3$ processes as they are
formally higher order. Furthermore, the essential ingredient in our
phenomenological analysis below is the lepton charge asymmetry
due to a non-vanishing $\lambda'_{i31}$ coupling. The
$2\ra3$ processes do not contribute, as their initial states are
charge symmetric and their contributions to the hadronic cross section
are only $5 \%$.

\mymed 

We end this section by presenting in Table~\ref{numericalresults}
selected cross section predictions for slepton production with $m_{\tl
i}= 100\gev$, $m_{\tl i}= 250\gev$ and $m_{\tl i}= 800\gev$ at the
Tevatron and the LHC via $\lam'_{i31}=0.1$.

\begin{table}[tb]
\begin{ruledtabular}
\begin{tabular}{c|c c}
 & Tevatron & LHC \\
\hline
\hline
$m_{\tl i} = 100 \,\text{GeV}$  & $25.5 $ fb & $3180 \,\, (2620)$ fb \\
$m_{\tl i} = 250 \,\text{GeV}$  & $2.10 \times 10^{-1}$ fb & $259\,\, (80.0)$ fb \\
$m_{\tl i} = 800 \,\text{GeV}$  & $2.86 \times 10^{-5}$ fb &  $11.6 \,\, (2.54)$ fb \\
\end{tabular}
\caption{\label{numericalresults} Hadronic Cross section predictions for 
$\tilde{\ell}_{i}^- t$ ($\tilde{\ell}_{i}^+ t$) production via $\lam'
_{i31}=0.1$ at the Tevatron ($\sqrt{S}=1.96$ GeV) and the LHC ($\sqrt{S}
=14$ GeV). Results are presented for the CTEQ6L1 \cite{Pumplin:2005rh} 
PDF parametrisation.}
\end{ruledtabular}
\end{table}

\section{Possible LHC Signatures}
\label{events}

Apart from the B$_3$ process, the sleptons and sneutrinos can decay
through gauge interactions.  Neglecting mixing between left- and
right-handed sleptons the possible tree-level decays are:
\begin{eqnarray}
{\tilde\ell}_i^- \ra
\left\{ \begin{array}{l}
\bar t\, d_k  \\[1mm]
\ell^-_i \tilde\chi^0_m  \\[1mm]
\nu_i \tilde\chi^-_n 
\end{array}
\right. \!\!,\qquad 
{\tilde\nu}_i \ra
\left\{ \begin{array}{l}
\bar b\, d_k   \\[1mm]
\nu_i \tilde\chi^0_m    \\[1mm]
\ell^-_i \tilde\chi^+_n   
\end{array}
\right.  \!. 
\label{slepton_decays}
\end{eqnarray}
The branching ratios depend on the masses of the sparticles, the
admixtures of the gauginos and on the size of the $\lam'_ {i3k}$
coupling.  We shall first assume, that the lightest neutralino,
$\tilde{\chi}_1^0$, is the lightest supersymmetric particle (LSP).
Possible decay modes via the $\lam'_ {i3k}$ interaction are: 
\begin{eqnarray}
\tilde{\chi}_1^0 \overset{\lam'}{\longrightarrow}
\left\{ \begin{array}{l}
\ell_i^+ \overline{t}\, d_k    \\[1mm]
\ell_i^- t\, \overline{d}_k   
\end{array}
\right.
,\qquad 
\tilde{\chi}_1^0 \overset{\lam'}{\longrightarrow}
\left\{ \begin{array}{l}
\bar \nu_i \bar b \,d_k  \\[1mm]
\nu_i b\,\overline{d}_k    
\end{array}
\right. \!. \quad 
\label{neutralino_decay}
\end{eqnarray}
Here the branching ratios depend mainly on the admixture of the
lightest neutralino.  The heavier neutralinos $\tilde{\chi}_{2,3,4}^0$
and the charginos $\tilde{\chi}_{1,2}^+$ dominantly decay into lighter
gauginos via gauge interactions, as in the P$_6$-MSSM. We have
neglected the decay $\tilde{\chi}^0_1\ra\nu\gamma$ \cite{Dawson:1985vr}, which
is suppressed except for very light neutralino masses
\cite{light-neutralino}.

\mymed

In SUSY scenarios, where the slepton (sneutrino) mass is of the order
of a few hundred GeV, the slepton (sneutrino) will decay dominantly
into the lightest neutralino and a lepton (neutrino). However,
significant chargino decay modes are also possible, if they are
kinematically accessible. Furthermore, decay chains involving a top
quark in the final state are either phase-space suppressed or
kinematically forbidden, unless the slepton is very heavy.
This affects the neutralino decays
(\ref{neutralino_decay}) involving charged leptons. Therefore, the
dominant hadron collider signatures of single slepton production in
association with a top quark are
\begin{align}
\begin{split}
        g d_k \ra{\tilde \ell_i}^- t \ra \ell_i^-\tilde{\chi}^0_1\,t&\ra
        \left\{ 
                \begin{array}{c}
        \ell_i^- \,(\bar \nu_i \bar b d_k)\, [b W^+]  \\
        \ell_i^- \,(\nu_i b \bar d_k)\, [b W^+]  \\
                \end{array}
        \right.\!\!. 
		\label{l_decays}
\end{split}
\end{align}
In parentheses are the neutralino LSP decay products
(\ref{neutralino_decay}); the particles in brackets arise from the top
quark decay. As mentioned before, for $k=1$ there is an asymmetry
between the number of positively and negatively charged leptons $\ell_
i^\pm$ at the LHC.

\mymed

The dominant signatures for a resonantly produced single sneutrino
are 
\begin{align}
\begin{split}
        \bar b\, d_k \ra{\tilde \nu_i} \ra
        \left\{ 
                \begin{array}{l}
        \bar b \,d_k  \\        
        \nu_i  \,(\bar \nu_i \bar b\, d_k)  \\
        \nu_i \, (\nu_i b \,\bar d_k)  \\
                \end{array}
        \right.\,,
		\label{nu_decays}
\end{split}
\end{align}
again the neutralino decay products are in parentheses. Although
the sneutrino production cross section at the LHC
(Fig.~\ref{xsection_sneutrino_LHC}) is up to two orders of magnitude
larger than the slepton plus top quark cross section
(Fig.~\ref{slepitop_xsection_LHC}), the event signature
(\ref{nu_decays}) is much harder to extract above the SM background.
It involves only two jets and possibly some missing transverse
energy. It therefore suffers from a large QCD background. However, if
the sneutrino decays into charginos and heavier neutralinos are
possible (\ref{slepton_decays}), we can have (additional) charged
leptons in the final state.

\mymed

We now consider SUSY scenarios, where the scalar tau (stau) is the LSP
instead of the lightest neutralino 
\cite{Allanach:2003eb,Allanach:2006st,Allanach:2007vi}. In this scenario
the lightest neutralino dominantly decays into a tau and the stau LSP,
$\tilde\chi_1^0\ra\stau_1^\pm \tau^\mp$. For $i=1,2$, the stau will dominantly
decay into a tau and a
virtual neutralino, leading to a four-body decay of the stau LSP
\cite{Allanach:2006st}.  The signatures for a stau LSP are
\begin{align}
\begin{split}
        g d_k \ra{\tilde \ell_i}^- t &\ra
        \left\{ 
                \begin{array}{l}
        \ell_i^-  \tau^\pm \,(\tau^\mp \bar \nu_i \bar b d_k)\, [b W^+]  \\[2mm]
        \ell_i^-  \tau^\pm \,(\tau^\mp \nu_i b \bar d_k) \,[b W^+] 
                \end{array}
        \right.\!. 
		\label{l_stau_decays}
\end{split}
\end{align}
The particles in parentheses are now the stau LSP decay products and
the particles in brackets are from the top quark decay. The difference
between the final states in Eqs.~(\ref{l_stau_decays}) and
(\ref{l_decays}) is, that for a stau LSP, the event is accompanied by
an additional pair of taus compared to scenarios with a neutralino
LSP. We find the same behaviour for the sneutrino decay chains.  It is
therefore easier to distinguish the signal from the background in stau
LSP scenarios as long as one is able to reconstruct the tau pair in
the final state.

\mymed

Note that for $i=3$ the two-body stau decay is kinematically
suppressed, or forbidden, due to the large top quark mass. The stau
LSP will in this case decay via a virtual top quark. Furthermore, we
can produce heavy staus, $\tilde{\tau}_2$, as well as light staus,
$\tilde{\tau}_1$, due to left-right mixing in the stau sector. In this case
the signatures are
\begin{align}
\begin{split}
        g d_k \ra{\tilde \tau_2}^- t &\ra
        \left\{ 
                \begin{array}{l}
        \tau^-  \tau^+  \,(\bar b d_k W^-)\, [b W^+]  \\[2mm]
        \tau^-  \tau^-  \,(b \bar d_k W^+) \,[b W^+]  \\[2mm]
        Z^0/h^0  \,(\bar b d_k W^-)\, [b W^+]
                \end{array}
        \right.\!,
		\label{stau2_decays}
\end{split}
\end{align}
and 
\begin{align}
\begin{split}
        g d_k \ra{\tilde \tau_1}^- t &\ra
        (\bar b d_k W^-)\, [b W^+] \, .
		\label{stau1_decays}
\end{split}
\end{align}
The particles in parentheses are the stau LSP decay products and those
in brackets are from the top quark decay. We see in
Eq.~(\ref{stau2_decays}) that one of the $\tilde{\tau}_2$
decay chains involves like-sign tau events. This can help to
distinguish signal from background although poor tau identification
could limit this possibility.

\section{Numerical Study for $\mathbf{\lam}_{231}^\prime\not=0$ 
and a $\tilde{\chi}^0_1$-LSP}
\label{details}

\subsection{The Scenario and Basic Cuts}

We now perform an explicit numerical study of single associated
slepton production. We focus on the more difficult case of a
neutralino LSP and restrict ourselves to $\lam'_{231}\not=0$, as
the dominant B$_3$ coupling. We assume that similar results can be
obtained for $\lam'_{131}\not=0$. A central analysis criterion will
be the lepton charge asymmetry of the final state.

\mymed 

According to Eq.~(\ref{l_decays}), the final-state signature to
examine is
\beq
\tilde\ell_L^\mp + t \;\longrightarrow\; \ell^\mp +( b + d + \nu) 
+[ b + W^\pm]\;,
\eeq
with the $W^\pm$ decaying hadronically. We thus have one charged
lepton, some missing $p_T$, and five jets, where two are b-quark jets. 
In our specific scenario, the charged lepton is a muon.

\medskip

The main background for this process is $t\bar t+j$ production
(which has recently been calculated at NLO \cite{Dittmaier:2007wz}) 
followed by the semi-leptonic decay of one of the top quarks. The second
background we examine is $b\bar b +W^\pm+{\rm jets}$ production followed by
the leptonic decay of the $W$ boson.

\mymed

\begin{table}[t]
  \begin{ruledtabular}
    \begin{tabular}{c||c|c|c|c|c}
      $\lamp_{231}$ & 0 & 0.1 & 0.2 & 0.3 & 0.4\\
      \hline
      $\mathrm{Br}(\tilde\mu^-\ra \bar t +d)$  & $0.0\%$ & $2.2\%$ & $8.4\%$ & $17.1\%$ & $26.8\%$\\
      $\mathrm{Br}(\tilde\mu^-\ra \mu^-+\tilde{\chi}^0_1)$  & $90.9\%$ & $88.9\%$ & $83.3\%$ & $75.4\%$ & $66.5\%$\\
      $\mathrm{Br}(\tilde\mu^-\ra \mu^-+\tilde{\chi}^0_2)$  & $3.2\%$ & $3.1\%$ & $2.9\%$ & $2.6\%$ & $2.3\%$\\
      $\mathrm{Br}(\tilde\mu^-\ra \nu_\mu +\tilde{\chi}_1^-)$  & $5.9\%$ & $5.8\%$ & $5.4\%$ & $4.9\%$ & $4.3\%$\\
    \end{tabular}
    \caption{\label{BRs_SPS1ap} Relevant branching ratios for ${\rm SPS1a}^\prime$
	for different couplings $\lamp_{231}$. }
  \end{ruledtabular}
\end{table} 

For our simulation, we assume an ${\rm SPS1a}^\prime$ similar scenario
\cite{AguilarSaavedra:2005pw}. We take the ${\rm SPS1a}^\prime$ spectrum and
couplings and add one B$_3$ coupling, $\lam'_{231}$.  The
relevant ${\rm SPS1a}^\prime$ masses are:
\begin{subequations}
\begin{eqnarray}
m_{\tilde\ell_L^\pm}&=&190\gev;\quad m_{\tilde\nu_\mu}=173\gev;\\
m_{\tilde{\chi}^0_1}&=&\phantom{1}98\gev; \quad m_{\tilde{\chi}^0_2}=184\gev;\;\;\;\phantom{.} \\
m_{\tilde{\chi}_1^\pm}&=&183\gev. 
\end{eqnarray}
\end{subequations}
All the charged slepton decays of
Eq.~(\ref{slepton_decays}) are therefore kinematically possible. The
corresponding branching ratios are given in Table \ref{BRs_SPS1ap} for
various couplings $\lamp_{231}$. Note that kinematically the sneutrino
can only decay via the neutralino or via the $\lamp_{231}$ coupling. The potential
signature would then be two jets possibly with some missing energy,
\textit{cf}. Eq.~(\ref{nu_decays}). 

\mymed

For the simulation of the single slepton plus top quark signal we have
written our own Monte Carlo program using the Les Houches
accord~\cite{Boos:2001cv} and linked this to \texttt{HERWIG6.5}
\cite{Corcella:2000bw,Corcella:2002jc,foot3}. The averaging of the
colour flow in the $s$- and $t$-channel single slepton production
diagrams is implemented by the method developed in
Ref.~\cite{Odagiri:1998ep}.  The supersymmetric particle spectra are
produced with \texttt{SOFTSUSY} \cite{softsusy}. The $t\bar
t$-background is simulated using the MC@NLO program
\cite{Frixione:2006gn,Frixione:2003ei}. The $b\bar b +W^\pm+{\rm jets}$
background is simulated by using \texttt{MadEvent}~\cite{Maltoni:2002qb}
to generate a sample of $b\bar b +W^\pm+2\ {\rm jet}$ events which are
then showered and hadronized using \texttt{HERWIG6.5}.
We use the CTEQ61 parton
distribution functions \cite{Pumplin:2005rh}. The top quark mass is
set to $m_t = 175 \gev$. 

\begin{figure}[ht!] 
  \setlength{\unitlength}{1cm} \includegraphics[scale=0.5]{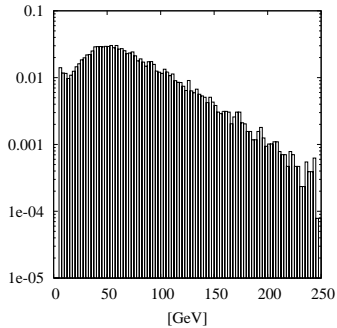}
        \caption{\label{fig:ptLepton}Relative $p_T$ distribution of
        the final-state signal $\ell^\pm$ at the LHC for ${\rm SPS1a}^\prime$,
		employing only the isolation cut on the lepton.}
\end{figure}

\begin{figure}[ht!] 
  \setlength{\unitlength}{1cm}
  \includegraphics[scale=0.5]{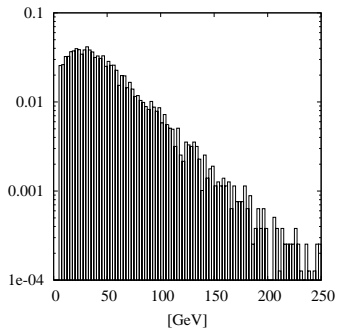}
  \caption{\label{fig:ptTtbarLepton}Relative $p_T$ distribution of the
  final-state $\ell^\pm$ from $t \bar t +j$ background at the LHC,
  employing only the isolation cut on the lepton.}
\end{figure}

Since our signature is very similar to the final state and
distributions of $t \bar t+j$ production followed by the semileptonic
decay, we use the standard set of CMS cuts for $t \bar t$ production
followed by the semileptonic decay, given in Ref.~\cite{CMSTDR} and
require an additional jet. This set of cuts leaves the large
semileptonic $t\bar t+j $ production, for which the cuts are designed,
and fewer $b\bar b +W^\pm+{\rm jets}$ events as backgrounds for the signal
process. The precise cuts are summarised below.

\mymed

The main difference between the semileptonically decaying top pair and
our signal is the $p_T$ distribution of the lepton steming from the
slepton compared to the one from the $W^\pm$ from one of the top
decays. We therefore compare in Figs.~\ref{fig:ptLepton} and
Fig.~\ref{fig:ptTtbarLepton} the $p_T$ distributions of the leptons
arising from the signal and the $t\bar t+j$ background processes,
respectively. We see, that the $p_T$ of the signal leptons has a peak
around 50 GeV. This peak corresponds to the mass difference between
the slepton and the neutralino with the energy carried away by the
lepton subtracted. The background lepton distribution peaks at 25 GeV
and then falls more steeply than the signal distribution for
increasing $p_T$. We thus harden the CMS semileptonic $t \bar t$ cut
for the isolated observed lepton from $p_T\ge20\gev$ to $p_T\ge35\gev
$.

\mymed

In addition to the charged lepton in the final state, we require two
tagged $b$-jets, as well as two further jets. Thus, the employed cuts
are
\begin{itemize}
\item 1 isolated lepton with pseudo-rapidity $\eta < 2.4$, $p_T > 35 \gev$. The isolation cut required less than $2\,$GeV of transverse energy in a cone of radius 0.4 around
the lepton direction.
\item 2 isolated $b$ jets and 2 non-$b$ jets, pseudo-rapidity $\eta < 2.4$,
  $p_T > 30\gev$.
\end{itemize}
The jets are defined using \texttt{PXCONE}~\cite{pxcone} which
uses the mid-point between two particles as a seed in addition to the particles
themselves to improve the infrared behaviour of the algorithm. A cone 
radius of 0.5 was used to define the jets. For the bottom and charm quarks
produced in the perturbative stage of the event the nearest jet in
$(\eta,\phi)$ is considered to have been produced by that quark if the distance
in $(\eta,\phi)$ was less than 0.2. 
We employ a $b$-tagging probability of $0.6$ and the probability for
mistagging a $c$-quark or light quark as a $b$-quark of $0.05$ and $0.02$ respectively.

\mymed

For the signal, we simulated $10^5$ events. Employing all cuts,
including $p_T(\ell^\pm)\geq\!35\gev$, we have $5\times10^3$ surviving
$\ell^-$ events and $1.7\times10^3$ surviving $\ell^+$ events. For the
$b\bar b +W^\pm+{\rm jets}$ background we simulated $10^6$ events for
both $W^-$ and $W^+$ production. After all cuts we are left
with $2.9\times10^4$ $\ell^-$ and $3.0\times10^4$ $\ell^+$ events,
respectively. $10^7$ $t\bar t+j$ events were simulated
resulting in $1.35\times10^5$ events for $\ell^-$ production and
$1.36\times10^5$ events for $\ell^+$production. 
This is summarised in Table~\ref{oldCutEvents}.

\mymed

For the simulated signal, we set $\lamp_{231}=0.053$.  In the
following we will estimate the signal for other values of $\lamp_
{231}$ by taking into account the $\lam'^2_{231}$ dependence of the
cross section. We also employ the $\lamp_{231}$ dependence of the the
$\tilde\mu^-\ra \mu^-+\tilde{\chi}^0_1$ branching ratio.

\mymed

\begin{table*}[tb]
  \begin{ruledtabular}
    \begin{tabular}{c||c|c|c|c}
      & simulated & $\ell^-$--events after cuts & $\ell^+$--events after cuts & Events $\cdot\mathrm{pb}^{-1}$\\
      \hline
      \hline
      signal  & 99\,900  & 5\,042 & 1\,664 & 0.0108  \\
      \hline
      $W^-\!+b\bar b+{\rm jets}$ bg\;  & 994\,000 & 28\,600 & 0 & 0.0431\\
      \hline
      $W^+\!+b\bar b+{\rm jets}$ bg\;\;  & 993\,500 & 0 & 29\,700 & 0.0625\\
      \hline
      $t\bar t+1\,$j bg\;& 9\,990\,500 & 135\,330 & 136\,360 & 22.00
    \end{tabular}
    \caption{\label{oldCutEvents} Results of simulating ${\rm SPS1a}^\prime$ with
      cuts given in the text. The
      number of leptons and the expected event rates are after
      cuts. }
  \end{ruledtabular}
\end{table*}

\mymed

\subsection{Lepton Charge Asymmetry}

In order to distinguish the signal from the background at the LHC
after these cuts, we propose as the decisive observable the lepton
charge asymmetry
\begin{equation}
\mathcal{A}_{\ell\pm}\equiv\frac{N_{\ell+}-N_{\ell-}}{N_{\ell+}+N_{\ell-}}
\; .
\end{equation}
Here $N_{\ell+}$ and $N_{\ell-}$ are the number of events with a
positively or negatively charged lepton, respectively. In
Fig.~\ref{slepitop_xsection_LHC}, we can see the separate signal cross
sections for $\ell^+$ and $\ell^- $ production at the LHC. For $m_{
\tilde\ell_L ^\pm}> m_t -m_d$, the $\ell^-$ cross section is 
significantly larger. This is due to the fact that the $d$-quark PDF
luminosity is significantly larger than that of the $\bar d$-quark for
$x\gsim10^{-2}$.

\mymed

We would expect the lepton charge asymmetry to be zero for the $t\bar
t+j$ background, as we have an equal number of top quarks and anti-top
quarks. For the background process $b\bar b +W^\pm+{\rm jets}$, we expect a
positive asymmetry, since the (valence) $u$-quark luminosity is
significantly larger than the (valence) $d$-quark luminosity in the
proton. For the signal, as we have seen, we expect a negative
asymmetry.

\mymed

However, in general, inclusive $t\bar t$ production has a charge
asymmetry in the final state at the LHC. It has been shown to be in
the range $[-0.025\%;0]$, if the detector has a symmetric acceptance
in the rapidity range $[-y_0;y_0]$.  For $y_0\ra \infty\;(0)$ the
asymmetry goes to $0\; (-0.025\%)$
\cite{Kuhn:1998kw,Bowen:2005xq}. This stems from the asymmetry in
$q\bar q$ induced $t\bar t$ production, which in turn is due to the
interference of C-odd and C-even modes, where C is the charge
conjugation operator. In the following, we will neglect this small
asymmetry because the statistical fluctuations lead to an even larger
asymmetry. The number of $\ell^\pm$ events in Table~\ref{oldCutEvents}
for the $t\bar t+\,$j background are consistent with a lepton charge
asymmetry of zero within two sigma.

\mymed

\begin{figure}[ht!]    
  \setlength{\unitlength}{1cm} 
\includegraphics[scale=0.38, bb = 60 77
  740 530, clip=true]{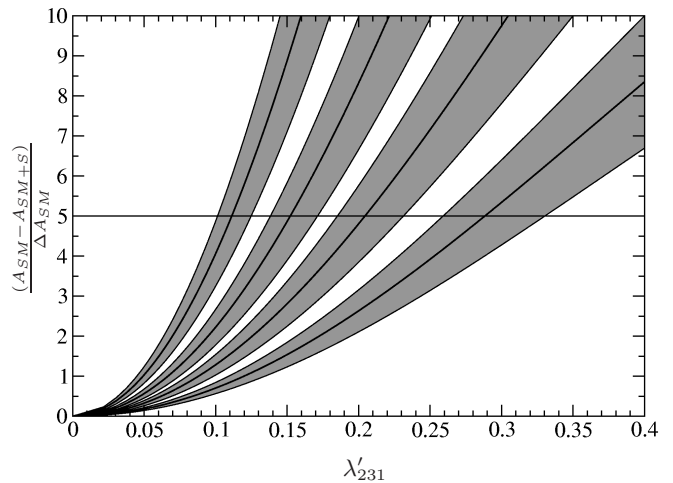}\put(-9.4,1.9){\rotatebox{90}
  {$\frac{(A_{SM}-A_{SM+S})}{\Delta A_{SM}}$}}
  \put(-5.0,-0.5){$\lamp_{231}$} 
\caption{\label{significance_SPS1a}
  Significance at the LHC as a function of $\lamp_{231}$ for ${\rm SPS1a}^\prime$
  with lepton $p_T \geq 35$ GeV. We show the significance for an
  integrated luminosity of 30$\,\mathrm{fb}^{-1}$ (lower curve),
  100$\,\mathrm{fb}^{-1}$, 300$\,\mathrm{fb}^{-1}$, and
  $1000\,\mathrm{fb}^{-1}$, respectively. Furthermore, we varied the
  signal cross section by $\pm 20 \%$ (grey region).}
\end{figure}

In Fig.~\ref{significance_SPS1a}, we show the significance, $\Sigma$,
of the signal for the ${\rm SPS1a}^\prime$ spectrum as a function of $\lam'_{231}$,
where
\begin{equation}
\Sigma\equiv \frac{(A_{SM}-A_{SM+S})}{\Delta A_{SM}}\,.
\end{equation}
Here $A_{SM}$ is the SM lepton charge asymmetry. $A_ {SM+S}$ is the
asymmetry for the signal and the SM background combined. $\Delta
A_{SM}$ is the error of the SM asymmetry prediction assuming purely
statistical errors for the number of positive and negative charged
leptons for each process separately, \textit{i.e.} $\sqrt{N_{\ell+}}$
and $\sqrt{N_{\ell-}}$.  The significance is shown for integrated
luminosities at the LHC of 30$\,\mathrm{fb}^{-1}$ (lower curves), 100$
\,\mathrm {fb}^{-1}$, 300$\,\mathrm{fb}^{-1}$, and 1000$\,\mathrm{fb}
^{-1}$, respectively.  We vary the cross section by $\pm 20\%$ (grey
region) to show possible effects due to higher order corrections for
the signal (\textit{cf.} Fig.~\ref{scale_LHC}).

\mymed

In Fig.~\ref{significance_SPS1a}, we see that for 30 fb$^{-1}$ we can
probe couplings down to about 0.3 for the ${\rm SPS1a}^\prime$
spectrum. In the ${\rm SPS1a}^\prime$ spectrum the squark mass is 544 GeV,
thus the experimental bound is $\lam'_{231}<1.0$,
\textit{cf.} Table~\ref{lamp_bounds}. For 300 fb$^{-1}$ we can probe 
couplings down to about 0.15. In the extreme case of 1000 fb$^{-1}$
this improves to about $\lam'_{231}=0.1$.

\medskip

\begin{table}[t]
  \begin{ruledtabular}
    \begin{tabular}{c||c|c|c|c|c}
      $\lamp_{231}$ & 0 & 0.1 & 0.2 & 0.3 & 0.4\\
      \hline
      $\mathrm{Br}(\tilde\mu^-\ra \bar t +d)$  & $0.0\%$ & $22.0\%$ & $53.0\%$ & $71.8\%$ & $81.9\%$ \\
      $\mathrm{Br}(\tilde\mu^-\ra \mu^-+\tilde{\chi}^0_1)$  & $60.9\%$ & $47.5\%$ & $28.6\%$ & $17.2\%$ & $11.0\%$ \\
      $\mathrm{Br}(\tilde\mu^-\ra \mu^-+\tilde{\chi}^0_2)$  & $13.8\%$ & $10.8\%$ & $6.5\%$ & $3.9\%$ & $2.5\%$ \\
      $\mathrm{Br}(\tilde\mu^-\ra \nu_\mu +\tilde{\chi_1}^-)$  & $25.3\%$ & $19.7\%$ & $11.9\%$ & $7.1\%$ & $4.6\%$ \\
    \end{tabular}
    \caption{\label{BRs_SPS1b} Relevant branching ratios for SPS1b
	for different couplings $\lamp_{231}$. }
  \end{ruledtabular}
\end{table}

We have repeated the above analysis for the parameter set SPS1b 
\cite{Allanach:2002nj}. Here we have the following masses:
\begin{subequations}
\begin{eqnarray}
m_{\tilde\mu_L}&=& 342\gev;\quad m_{\tilde\nu_\mu}= 333\gev; \label{smuon-mass}\\
m_{\tilde\chi^0_1}&=& 163\gev;\quad m_{\tilde\chi^0_2}= 306\gev;\;\;\;\phantom{.}\\
m_{\tilde\chi^\pm_1}&=& 306\gev\,. 
\end{eqnarray}
\end{subequations}
We show the branching ratios for different $\lamp_{231}$ in Table
\ref{BRs_SPS1b}. We see, that the B$_3$ decay into a d quark and a 
top quark is the dominant decay for large $\lamp_{231} $,
\textit{i.e.}  $\lamp_{231}>0.19$. One might thus consider an analysis
based on this decay mode. However the signature is $t\bar t+\,j$,
which has a very large background. We thus continue to consider the
neutralino decay mode.  The significance will then approach a constant
value for a constant luminosity and large $\lamp_{231}$, because the
cross section and the B$_3$ decay both scale with $\lam'^2_
{231}$. Furthermore the slepton mass is now significantly larger, but
so is the lightest neutralino mass. The mass difference however has
grown, leading to significantly higher charged lepton $p_T$'s
compared to ${\rm SPS1a}^\prime$,
\textit{cf} Fig.~\ref{fig:ptLepton}. We thus impose the stricter
cut on the lepton transverse momentum
\begin{equation}
p_T(\ell^\pm)\geq 70\gev\,.
\end{equation}

\begin{figure}[ht!] 
  \setlength{\unitlength}{1cm} \includegraphics[scale=0.38, bb = 60 77
  740 530, clip=true]{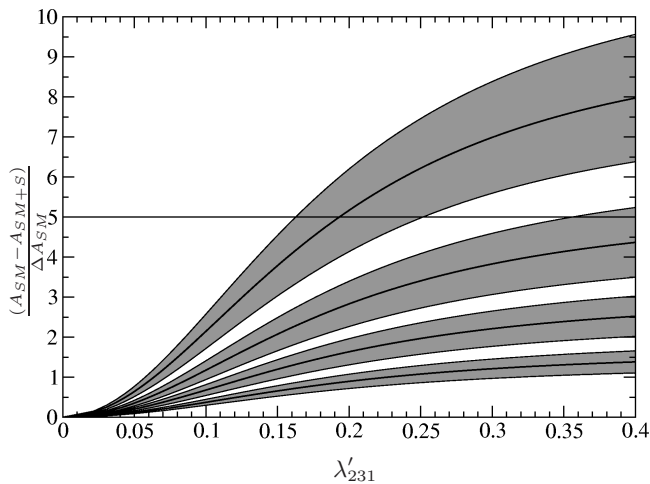}
	\put(-9.3,1.6){\rotatebox{90}{$\frac{(A_{SM}-A_{SM+S})}{\Delta A_{SM}}$}}
	\put(-5.0,-0.5){$\lamp_{231}$}
\caption{\label{significance_SPS1b} Same as for
          Fig.~\ref{significance_SPS1a}, but for the parameter set SPS1b
		  with lepton $p_T \geq 70$ GeV.}
\end{figure}

The results are shown in Fig.~\ref{significance_SPS1b}. In this case,
for the relatively low integrated luminosity of 30 fb$^{-1}$ we have
no chance of observing the signal via the lepton asymmetry; the
neutralino branching fraction is too small to have enough events. In
fact, it is only for the extremely high integrated luminosity of 1000
fb$^{-1}$ that we have a significant sensitivity range, down to about
$\lam'_{231}=0.2$.

\medskip

\begin{table}[t]
  \begin{ruledtabular}
    \begin{tabular}{c||c|c|c|c|c}
      $\lamp_{231}$ & 0 & 0.1 & 0.2 & 0.3 & 0.4\\
      \hline
      $\mathrm{Br}(\tilde\mu^-\ra \bar t +d)$  & $0.0\%$ & $23.7\%$ & $55.4\%$ & $73.6\%$ & $83.2\%$\\
      $\mathrm{Br}(\tilde\mu^-\ra \mu^-+\tilde{\chi}^0_1)$  & $100\%$ & $76.3\%$ & $44.6\%$ & $26.4\%$& $16.8\%$ 
    \end{tabular}
    \caption{\label{BRs_SPS1bEXTREME} Relevant branching ratios for 
	the high $p_T$ scenario for different couplings $\lamp_{231}$. 
	The scenario is described in the text.}
  \end{ruledtabular}
\end{table}

\begin{figure}[ht!] 
  \setlength{\unitlength}{1cm} \includegraphics[scale=0.38, bb = 60 75
  740 530, clip=true]{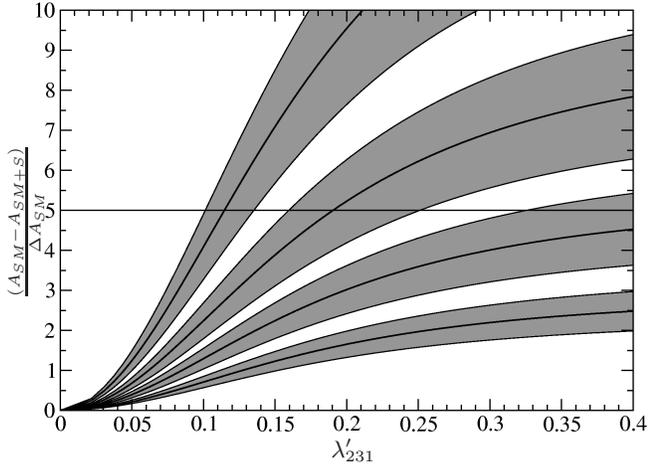}
  \put(-9.3,1.8){\rotatebox{90}{$\frac{(A_{SM}-A_{SM+S})}{\Delta
  A_{SM}}$}} \put(-5.0,-0.3){$\lamp_{231}$}
  \caption{\label{significanceSPS1bEXTREME} Same as for
  Fig.~\ref{significance_SPS1a}, but for the high-$p_T$ scenario
  and with lepton $p_T \geq 120$ GeV. The scenario is described
  in the text.}
\end{figure}

In order to see what can be probed at the LHC, we have chosen as a
third example a mass spectrum which optimizes our signal. For this we
considered a modified SPS1b spectrum, where we first lowered the mass
of the lightest neutralino to
\begin{eqnarray}
m_{\tilde\chi^0_1}&=& 80\gev\,,
\end{eqnarray}
in order to obtain a larger mass difference between the smuon and the
lightest neutralino. We can then harden the $p_T$ cut to
\begin{equation}
p_T(\ell^\pm)\geq 120\gev\,.
\end{equation}
This leads to a better signal to background ratio compared to SPS1b.
Second, we increased the masses of $\tilde{\chi}^0_2$ and $\tilde{\chi}^\pm_1$ to
\begin{eqnarray}
m_{\tilde\chi^0_2}=m_{\tilde\chi^\pm_1}&=& 450\gev\,.
\end{eqnarray} 
This increases the $\tilde\mu^-\ra \mu^-+\tilde{\chi}^0_1$ branching ratio
compared to SPS1b, because decays into heavier neutralinos and into
charginos are now kinematically forbidden.  We show the relevant
branching ratios for different $\lamp_{231}$ in Table
\ref{BRs_SPS1bEXTREME}.  We refer to this scenario as the high-$p_T$
scenario.  The resulting significance for the high-$p_T$ scenario is
shown in Fig.~\ref{significanceSPS1bEXTREME}.

As can be seen, for an integrated luminosity of 30 fb$^{-1}$ we still
have no sensitivity in $\lam'_{231}$. But now for an integrated
luminosity of 300 fb$^{-1}$, we can probe couplings down to 0.19, well
below the experimental bound of 1.5, \textit{cf}. Table 
\ref{lamp_bounds} where now $m_{\tilde{b}_L}=830$ GeV in SPS1b. 
For an integrated 
luminosity of 1000 fb$^{-1}$ we can probe couplings down to 0.11.

\mymed

The influence of systematic errors in the background cross section on
our sensitivity are small. Varying the $t\bar t+j$ cross section by
$+10\%$ ($-10\%$) changes the asymmetry by roughly $-9\%$ ($+11\%$).
Varying the $b\bar b +W^\pm+{\rm jets}$ cross section by $\pm10\%$ only
effects the asymmetry by $\mp1.6\%$ for ${\rm SPS1a}^\prime$ with $\lamp_{231}=0.3$
and by $\mp1.2\%$ for the high-$p_T$ scenario with $\lamp_
{231}=0.3$. Yet, detector effects resulting in an error on the
observed charge asymmetry are a problem. Misalignment in the detector
can lead to a difference in $p_T$ measurement of positive and negative
leptons, respectively. This will lead to an observed, effective charge
asymmetry after a cut on the lepton $p_T$ \cite{Blusk:2007zza}. An
analysis of this must be performed by the experimentalists and is well
beyond the scope of this paper.

\mymed

For ${\rm SPS1a}^\prime$ with $\lamp_{231}=0.3$, a simulated detector based charge
asymmetry of $0.66\%$ leads to an asymmetry of the $t \bar t$
background of the same size as that of the signal. For the special
case chosen with high $p_T$ leptons in the final state, \textit{i.e.} 
the high-$p_T$ scenario with $\lamp_{231}=0.3$, a simulated asymmetry
of $0.89\%$ would lead to the same effect. Therefore, a higher $p_T$
cut is less sensitive to systematic errors, due to the high $p_T$ cut
effecting the $t \bar t$ background.

\section{Conclusion}
\label{conclusion}
For the special case of dominant B$_3$ couplings $\lam'_{i3k}$
resonant charged slepton production is not possible at hadron
colliders, as there are no incoming top quarks in the proton. We must
then consider the associated production with a top quark. We have
analyzed this difficult signature in detail. As the decisive
observable for $\lam'_{131}$ and $\lam'_{231}$ 
we propose the lepton charge asymmetry. For the
supersymmetric spectrum ${\rm SPS1a}^\prime$ we have found a significant sensitivity
range at the LHC, summarized in Fig.~\ref{significance_SPS1a}. For the
heavier spectrum SPS1b the LHC is significantly less sensitive 
in the coupling, as can be seen in
Fig.~\ref{significance_SPS1b}. We have constructed a heavy spectrum
with a larger slepton-neutralino mass difference in order to explore
the maximum sensitivity at the LHC. This is shown in 
Fig.~\ref{significanceSPS1bEXTREME}.

\begin{acknowledgments}
  We thank Sebastian Fleischmann, Nicolas M\"oser and Jan Schumacher
  for helpful discussions about the ATLAS detector. We thank Olaf Kittel
  for reading parts of the manuscript. MB thanks the
  IPPP, Durham, for warm hospitality offered during various stages of
  this paper. SG thanks the Deutsche Telekom Stiftung for financial
  support. MB and HD thank the BMBF grant 05 HT6PDA `Rekonstruktion
  von Parametern der supersymmetrischen Erweiterungen des
  Standardmodells am LHC' for financial support.
\end{acknowledgments}

\bibliographystyle{h-physrev}


\begin{thebibliography}{99}
\bibitem{foot3}
The version of HERWIG used in this paper includes modifications to
simulate the four-body decays of the stau LSP and is available on
request from Peter Richardson.
\end{thebibliography}

\end{document}